\documentclass[fleqn,twoside]{article}
\usepackage{espcrc2}

\usepackage{graphicx}
\usepackage[figuresright]{rotating}

\newcommand{\AmS}{{\protect\the\textfont2
  A\kern-.1667em\lower.5ex\hbox{M}\kern-.125emS}}

\title{OPERA-CNGS/ Fr\'ejus-SPL}

\author{J. E. Campagne\address{Laboratoire de L'acc\'el\'erateur Lin\'eaire, Universit\'e Paris Sud, \\ BP 34, 91898 Orsay Cedex},
        A. Cazes\addressmark\thanks{contact person, mail : cazes@lal.in2p3.fr}}

\begin{document}


\maketitle

\section{Introduction}

The poster presented the OPERA experiment and the SPL-Fr\'ejus neutrino super beam project.

OPERA \cite{dario} is an experiment willing to see tau neutrino apperance. The detector is discribed in section \ref{sec:opera}  and the expected results are given.

Section \ref{sec:SPL} is deveted to the SPL neutrino super beam \cite{SPL}, wich search for $\theta_{13}$. An optimisation of the SPL energy is proposed.

\section{The OPERA experiment}
\label{sec:opera}

The experiment is located in the Gran Sasso Laboratory and recieves
the CNGS $\nu_\mu$ beam created at CERN, 732~km form Gran Sasso. OPERA
detectes the $\tau$ created by charged current interactions of
oscillated $\nu_\tau$. The target is made of 1800~t of lead, cut in
1~mm thick plates, piled together with emulsion films inside
bricks. The high space resolution of emulsion allows to identify the
$\tau$ decay topology. There are more than 200,000 such brick in
OPERA, arranged in walls. Hit bricks are localized using scintillator
strips installed between the brick walls and two spetrometers allow to
identify muons with their charge. This divides by a factor 20 the
charm backgroud (charmed mesons have a decay topology similar to the
tau, but they produce wrong charge muons).

If $\Delta m^2_{23}=2.4\times10^{-3}eV^2$, one expect 10.5 $\nu_\tau$
events identified in OPERA with an expected backgroud less than 0.7
event for 5 years running at $4.5\times10^{19}$pot/year.

\section{The SPL neutrino super beam}
\label{sec:SPL}

Neutrino super beams ($\nu_\mu$/$\bar{\nu}_\mu$) can be used for the
search of $\theta_{13}$ and $\delta_{CP}$. $\nu_e$
appearance may be seen for instance in a 440~kt water \v{Cerenkov}
detector located in a new cavity in the Fr\'ejus laboratory at 130~km
from CERN.


A 4~MW proton beam called Super Proton Linac (SPL) is under
study at CERN. Its protons imping a mercury target to
produce pions focalized by two concentric electromagnetic
horns. Their shape have been optimised to produce 260~MeV neutrinos
beam, which is the first oscillation maximum at $\Delta
m^2_{23}=2.5\times10^{-3}eV^2$.  The horns are followed by a 40~m
long, 2~m radius decay tunnel.

A full simulation of the beam line have been set up, including the target
station, the focusing horns and the decay tunnel geometry. Special
attention has been taken for the simulation of kaons background which
is crucial for the SPL energy optimisation study \cite{jecac}. It uses
algorithm based on the probability that have any neutrino of the
simulation to reach the detector.

Up to now, the nominal kinetic energy of the SPL is 2.2~GeV,
 but the result of \cite{jecac} indicates that an energy of 3.5~GeV could improve  the sensitivity by 20\%, allowing  to measure $\theta_{13}$ independetly of the value of $\delta_{CP}$ down to $\sin^22\theta_{13} = 2.02\times10^{-3}$, at 90\% CL.

\end{document}